\documentstyle[aps,prl,twocolumn,epsfig]{revtex}

\newcommand{\etal}{{\em et al.}}                %       et al.

\newcommand{\dn}[2]{d^{#1}{#2}\,}
\newcommand{\eqref}[1]{(\ref{#1})}

\begin{document}

\title{Pion Imaging at the AGS}

 \author{S.Y.~Panitkin,$^7$ N.N.~Ajitanand,$^{12}$ J.~Alexander,$^{12}$ M.~Anderson,$^5$
  D.~Best,$^1$ F.P.~Brady,$^5$ T.~Case,$^1$ W.~Caskey,$^5$
  D.~Cebra,$^5$ J.~Chance,$^5$ P.~Chung,$^{12}$
  B.~Cole,$^4$ K.~Crowe,$^1$ A.~Das,$^{10}$
  J.~Draper,$^5$ M.~Gilkes,$^{11}$ S.~Gushue,$^2$
  M.~Heffner,$^5$ A.~Hirsch,$^{11}$
  E.~Hjort,$^{11}$ L.~Huo,$^6$ M.~Justice,$^7$
  M.~Kaplan,$^3$ D.~Keane,$^7$ J. Kintner,$^8$
  J.~Klay,$^5$ D.~Krofcheck,$^9$ R.~Lacey,$^{12}$ J.~Lauret,$^{12}$
  M.A.~Lisa,$^{10}$ 
  H.~Liu,$^7$ Y.M.~Liu,$^6$ R.~McGrath,$^{12}$ Z.~Milosevich,$^3$,
  G.~Odyniec,$^1$ D.~Olson,$^1$  C.~Pinkenburg,$^{12}$
  N.~Porile,$^{11}$ G.~Rai,$^1$ H.G.~Ritter,$^1$
  J.~Romero,$^5$ R.~Scharenberg,$^{11}$ L.S.~Schroeder,$^1$
  B.~Srivastava,$^{11}$ N.T.B.~Stone,$^2$ T.J.M.~Symons,$^1$
  S.~Wang,$^7$ R.~Wells,$^{10}$ J.~Whitfield,$^3$ T.~Wienold,$^1$ R.~Witt,$^7$
  L.~Wood,$^5$ X.~Yang,$^4$ W.N.~Zhang,$^6$ Y.~Zhang $^4$\\
  (E895 Collaboration)
 }

 \address{
 $^1$Lawrence Berkeley National Laboratory, Berkeley, California 94720 \\
 $^2$Brookhaven National Laboratory, Upton, New York 11973 \\
 $^3$Carnegie Mellon University, Pittsburgh, Pennsylvania 15213 \\
 $^4$Columbia University, New York, New York 10027 \\
 $^5$University of California, Davis, California 95616 \\
 $^6$Harbin Institute of Technology, Harbin 150001, P.~R.~China \\
 $^7$Kent State University, Kent, Ohio 44242 \\
 $^8$St.~Mary's College of California, Moraga, California 94575 \\
 $^9$University of Auckland, Auckland, New Zealand \\
 $^{10}$The Ohio State University, Columbus, Ohio 43210 \\
 $^{11}$Purdue University, West Lafayette, Indiana 47907 \\
 $^{12}$State University of New York, Stony Brook, New York 11794 \\
 }

% typeset front matter
\maketitle

\begin{abstract}
Source imaging analysis was performed on two-pion
correlations in central Au + Au collisions at beam
energies between 2 and 8$A$ GeV.  We apply the imaging technique by
Brown and Danielewicz, which allows 
a model-independent extraction of
source functions with useful accuracy out to relative
pion separations of about 20 fm. We found that extracted source functions have
Gaussian shapes.
Values of source functions at zero separation are almost constant
across the energy range under study.
Imaging results are found to be
consistent with conventional source parameters obtained from a
multidimensional HBT analysis.      
\end{abstract}

\section{Introduction}
Measurements of two-particle correlations offer a powerful tool for
studies of spatial and temporal behaviour of the heavy-ion collisions.
The complex nature of the heavy-ion reactions requires utilization of
different probes and different analysis techniques in order to obtaine
reliable and complete picture of the system created in the
collision.\\
 We will discuss application of imaging technique of Brown-Danielewicz 
which allows one to reconstruct the entire source function, phase space
density and entropy from any like-pair correlations in a model 
independent way.
We will present recent results of the two-pion imaging analysis
performed by the E895 Collaboration and discuss natural connection
between imaging and traditional HBT. 
\subsection{E895 Experiment and Analysis Details}
Detailed description of the experiment E895 can be found
elsewhere~\cite{rai_93}. 
The Au beams of kinetic energy 1.85, 3.9, 5.9 and 7.9$A$ GeV from the
AGS accelerator 
at Brookhaven were incident on a fixed Au target.  The analyzed
$\pi^-$ samples come from the main E895 subsystem --- the EOS Time
Projection Chamber (TPC)~\cite{rai_90}, located in a dipole magnetic
field of 0.75 or 1.0 Tesla.   Results of the E895
$\pi^-$ HBT analysis were reported earlier~\cite{lisa_1}, and in this
paper, we focus on an application of imaging to the same datasets.

\subsection{Source Imaging}
In order to obtain the two-pion correlation function $C_2$
experimentally, the standard event-mixing technique was 
used~\cite{kopylov_74}. 
 Negative pions were detected and
reconstructed over a substantial fraction of $4\pi$ solid angle in
the center-of-mass frame, and simultaneous measurement of particle
momentum and specific ionization in the TPC gas helped to separate
$\pi^-$ from other negatively charge particles, such as electrons,
K$^-$ and antiprotons.  Contamination from these species is estimated
to be under 5\%, and is even less at the lower beam energies and
higher transverse momenta.  Momentum resolution in the region of 
correlation measurements is better than 3\%.
Event centrality selection was based on the multiplicity of 
reconstructed charged particles.  In the present analysis, events were
selected with a multiplicity corresponding to the upper 11\% of
the inelastic cross section for Au + Au collisions.  Only pions with 
$p_T$ between 100 MeV/$c$ and 300 MeV/$c$ and within $\pm 0.35$ units 
from midrapidity were used.
Another requirement was for each used $\pi^-$ track to point back to the primary 
event vertex with a distance of closest approach (DCA) less than 2.5 cm.
This cut removes most pions originating from weak decays of long-lived
particles, e.g. $\Lambda$ and K$^0$.  Monte Carlo simulations based on 
the RQMD model~\cite{sorge_95} indicate that decay daughters are present 
at a level that varies from 5\% at 4$A$ GeV to 10\% at 8$A$ GeV, and   
they lie preferentially at $p_T < 100$ MeV/$c$.  
Finally, in order to overcome effects of track merging, a cut on spatial
separation of two tracks was imposed.  For pairs from both ``true'' and 
``background'' distributions, the separation between two tracks was 
required to be greater than 4.5 cm over a distance of 18 cm in the beam 
direction. This cut also suppresses effects of track
splitting~\cite{lisa_1}. Correlation functions were corrected for pion
Coulomb interaction~\cite{lisa_1}. 
 \begin{figure}
\begin{center}	
 \epsfig{file=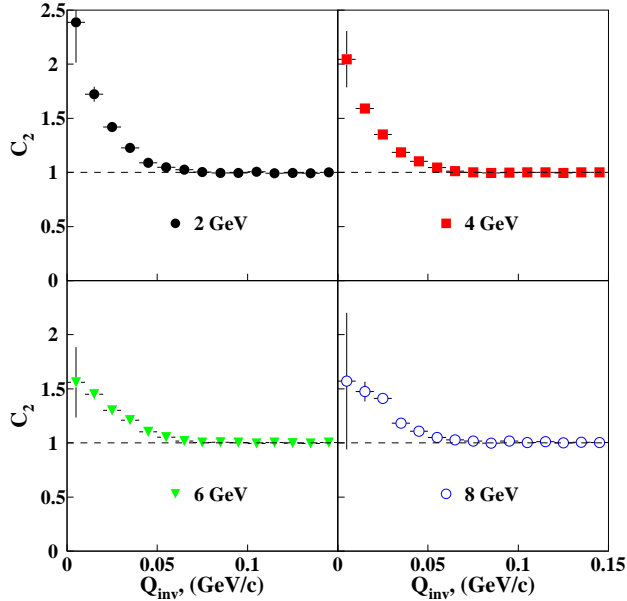,width=9.0cm}
 \caption{Measured two-pion correlation functions for Au + Au central
 collisions at four beam energies. }
 \label{panitkin_1}
\end{center}
 \end{figure}
Imaging~\cite{dbrown_1,dbrown_2} was used to extract quantitative
information from the measured correlation functions. It has been
shown~\cite{panitkin_99_1} that  
imaging allows robust reconstruction of the source function for systems 
with non-zero lifetime, even in the presence of strong space-momentum
correlations.  The main features of the method are outlined below; see 
Refs~\cite{dbrown_1,dbrown_2,dbrown_00_1} for more details.  The 
two-particle correlation function may be expressed as follows: 
\begin{equation}
        C_{\bf P}({\bf Q}) -1 =
        \int d{\bf r} \,K({\bf Q}, {\bf r}) \, S_{\bf P} ({\bf r}) \,,
        \label{panitkin_K}
\end{equation}
where $K = |\Phi_{\bf Q}^{(-)}({\bf r})|^2-1$ and $\Phi_{\bf Q}^{(-)}$ is the 
relative wavefunction of the pair.  The source function, $S_{\bf P} ({\bf r})$,
is the distribution of relative separations of emission points for the two 
particles in their center-of-mass frame.  The 
total momentum of the pair is denoted by ${\bf P}$, the relative momentum 
by ${\bf Q}={\bf p_1}-{\bf p_2}$, and the relative separation 
by ${\bf r}$.  Since the correlation data is already Coulomb corrected, we 
may approximate the relative wavefunction of the pions with a noninteracting
spin-0 boson wavefunction:
 \begin{figure}
 \begin{center}
 \epsfig{file=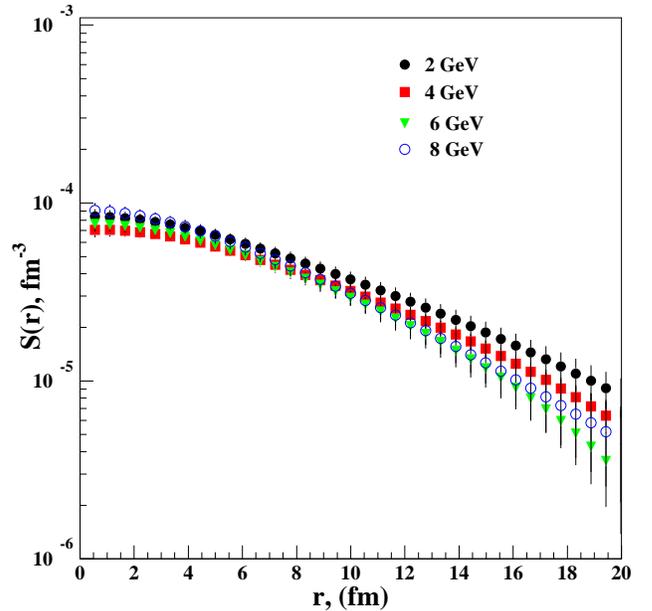,width=9.0cm}
 \caption{ Relative source functions extracted from the pion
 correlation data at 2, 4, 6 and 8$A$ GeV. }
 \label{panitkin_2}
 \end{center}
 \end{figure}
\begin{equation}
        % \[\wftn{q}{r}
        \Phi_{{\bf Q}}^{(-)}({\bf r})=\frac{1}{\sqrt{2}}\left
         ( e^{i{\bf Q}/2\cdot{\bf r}} +e^{-i {\bf Q}/2\cdot {\bf r}}\right).
        %   K_0 (q,r) =\frac{\sin{(2 q r)}}{2 q r}
\end{equation}
The goal of imaging is the determination of the relative
source function ($S_{\bf P} ({\bf r})$ in Eq.~(\ref{panitkin_K})), 
given $C_{\bf P}({\bf Q})$.  The problem of imaging then becomes the
problem of inverting $K({\bf q}, {\bf r})$~\cite{tarantola}.
For pions, one might think that because Eq.~\ref{panitkin_K} 
is a Fourier cosine transform it 
may be inverted analytically~\cite{dbrown_1} to give the source function
directly in terms of the correlation function:
\begin{table}
\caption{Radius parameters of Gaussian fits to the extracted source
($R_S$) functions and measured correlation functions($R_{C2}$) for
different energies.} 
\begin{tabular}{lcccc}
 $E_{b}$ (AGeV)  & 2 & 4 & 6 & 8 \\
 \hline
 $R_{S}$(fm)    &6.70$\pm$0.04& 6.35$\pm$0.03  & 5.56$\pm$0.03 & 5.53$\pm$0.05\\
 $R_{C2}$ (fm)  &6.39$\pm$0.2& 6.05$\pm$0.1  & 5.51$\pm$0.15  & 5.61$\pm$0.28 \\
\end{tabular}
\label{panitkin_t_1}
\end{table}
\begin{equation}
        S_{\bf p}({\bf r}) = \frac{1}{(2\pi)^3}\int\dn{3}{Q}
                     \cos{({\bf Q}\cdot{\bf r})}
                     \left(C_{\bf p}({\bf Q})-1\right).
\end{equation}
While this might work for vanishing experimental uncertainty, for 
realistic situations, this is a poor way to perform the inversion.  When 
Fourier transforming, it is impossible to distinguish between statistical noise
and real structure in the data \cite{FTisbad}.
Instead, we proceed as in \cite{dbrown_in_prep} and 
expand the source in a Basis Spline basis: $S({\bf r})=\sum_j S_j B_j({\bf r})$.
With this, Eq.~\eqref{panitkin_K} becomes a matrix equation
$C_i=\sum_j K_{ij} S_j$ with a new kernel:
\begin{equation}
        K_{ij}=\int d{\bf r} K({\bf Q}_i,{\bf r}) B_j({\bf r}).  
\end{equation}
Imaging reduces to finding the 
set of source coefficients, $S_j$, that minimize the $\chi^2$.  Here, 
$\chi^2=\sum_i(C_i - \sum_j K_{ij} S_j)^2/\Delta^2C_i$.\\
This set of source coefficients is 
$S_j=\sum_i[(K^T(\Delta^2C)^{-1}K)^{-1}K^TB]_{ji} (C_i-1)$ 
where $K^T$ is the transpose of the kernel matrix.
The uncertainty of the source is the square-root of the diagonal elements of the
covariance matrix of the source,
$\Delta^2S=(K^T(\Delta^2C)^{-1}K)^{-1}$.
It has been shown~\cite{dbrown_00_1} that for the case of noninteracting 
spin-zero bosons with Gaussian correlations, there is a natural connection 
between the imaged sources and ``standard'' HBT parameters.
Indeed, the source is a Gaussian with the ``standard'' HBT parameters as radii:
\begin{equation}
        S({\bf r})=\frac{\lambda}{(2\sqrt{\pi})^3\sqrt{\det{[R^2]}}}
        \exp{\left(-\frac{1}{4}r_ir_j[R^2]^{-1}_{ij}\right)}\,,
\label{s_gauss}
\end{equation}
where $\lambda$ is a fit parameter traditionally called the chaoticity 
or coherence factor in HBT analyses.  Here, $[R^2]$ is the real symmetric 
matrix of radius parameters 
\begin{equation}
 [R^2]=\left(
\begin{array}{ccc}
                 R_o^2 & R_{os}^2 & R_{o\ell}^2 \\
                 R_{os}^2 & R_s^2 & R_{s\ell}^2 \\
                 R_{o\ell}^2 & R_{s\ell}^2 & R_\ell^2
\end{array}
\right).
\end{equation}
Eq.~(\ref{s_gauss}) is the most general Gaussian one may use, but 
usually one assumes a particular symmetry of the single particle
source so that there is only one non-vanishing non-diagonal element 
$R_{ol}^2$.
The asymptotic value of the relative source function at zero 
separation $S({\bf r}\rightarrow 0)$ is related to the inverse 
effective volume of particle emission, and has units of fm$^{-3}$:
\begin{equation}
        S({\bf r}\rightarrow
        0)=\frac{\lambda}{(2\sqrt{\pi})^3\sqrt{\det{[R^2]}}} \,.
\label{connect}
\end{equation}
As it was shown in~\cite{dbrown_00_1} $S({\bf r}\rightarrow 0)$ is an
 important parameter needed to extract the space-averaged phase-space density. 
Figure~\ref{panitkin_1} shows measured angle-averaged 
two-pion correlation functions 
for central Au + Au collisions at 2, 4, 6 and 8$A$ GeV.
Figure~\ref{panitkin_2} shows relative source functions $S(r)$ 
obtained by applying the imaging technique to the measured two-pion 
correlation functions.    
Note that the plotted points are for representation of the continuous
source function and hence are not statistically 
independent of each other as the source functions are expanded 
in Basis Splines~\cite{dbrown_in_prep}. Since the source covariance
matrix is not diagonal, the coefficients of the Basis spline expansion are also
not independent which is taken into account during $\chi^2$
calculations.
 \begin{figure}
 \begin{center}
 \epsfig{file=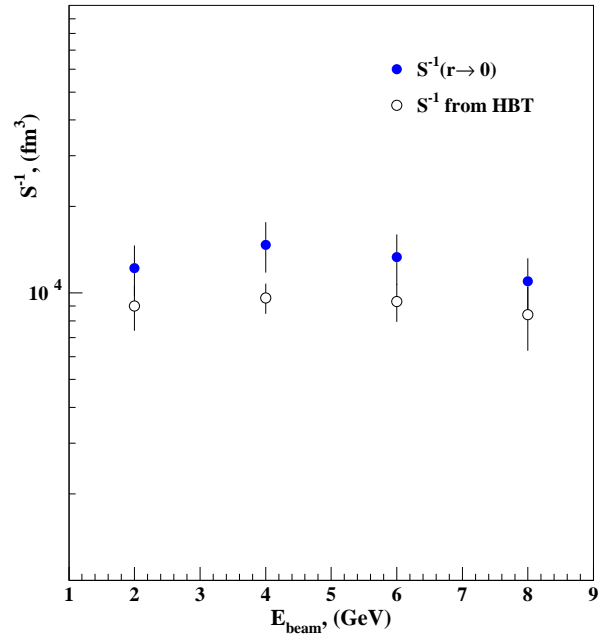,width=9.0cm}
 \caption{Values of inverse relative source functions at zero separation as a
 function of beam energy, obtained using imaging (solid circles) and HBT (open
circles).}
 \label{panitkin_3}
 \end{center}
 \end{figure}
\begin{table}
\caption{Fit parameters for the Bertsch-Pratt HBT
parameterization of the pion correlation functions for E895 beam
energies used for evaluation of $S(r\rightarrow 0)$.}
\begin{tabular}{lcccc}
 $E_{b}$ (AGeV)  & 2 & 4 & 6 & 8 \\
 \hline
 $\lambda$       &0.99$\pm$0.06& 0.74$\pm$0.03 & 0.65$\pm$0.03 & 0.65$\pm$0.05\\
 $R_o$  (fm)     &6.22$\pm$0.26& 5.79$\pm$0.16  & 5.76$\pm$0.23 &
5.49$\pm$0.31\\
 $R_{s}$ (fm)    &6.28$\pm$0.20& 5.37$\pm$0.11  & 5.05$\pm$0.12  & 4.83$\pm$0.21
\\
 $R_{l}$ (fm)    &5.15$\pm$0.19& 5.15$\pm$0.14  & 4.72$\pm$0.18 &
4.64$\pm$0.24\\
 $R_{ol}^2$ (fm)&-2.43$\pm$1.71& 0.43$\pm$1.03& 2.17$\pm$1.20 & -0.65$\pm$1.85
\\
\end{tabular}
\label{panitkin_t_2}
\end{table} 
With this technique, we may reconstruct the distribution of 
relative pion separations with useful accuracy out to $r \sim 20$ fm.
The images obtained at each of the four E895 beam energies are rather 
similar in shape, and upon fitting with a Gaussian function, values of 
$\chi^2$ per degree of freedom between 0.9 and 1.2 are obtained.  
Results of fits to the source function and correlation functions are
shown in Table~\ref{panitkin_t_1}. One can see that source radii
extracted via both techniques are similar, further
confirming the validity of a Gaussian source hypothesis.
Figure~\ref{panitkin_3} compares values related to effective volumes of
pion emission
inferred from the standard Bertsch-Pratt pion HBT parametrization 
(open circles) with the effective volumes derived from the image 
source functions shown in Fig.~\ref{panitkin_2} (solid circles).  
Essentially, this figure plots the inverse of the left- and right-hand 
sides of Eq.~\ref{connect}.
 Results of the multidimensional Bertsch-Pratt 
fit to the pion correlation data have already been published in 
Ref.~\cite{lisa_1} and are reproduced in Table~\ref{panitkin_t_2}.
It can be seen from Fig.~\ref{panitkin_3} that the agreement between
imaging and the HBT parametrization is fairly good.
Values of source functions at zero separations estimated via either technique
are approximately constant within errors across the 2 to 8$A$ GeV beam
energy range.\\
In summary, we present measurements of one-dimensional correlation functions 
for negative pions emitted at mid-rapidity from central Au + Au collisions 
at 2, 4, 6 and 8$A$ GeV.  These correlation functions are analyzed using the 
imaging technique of Brown and Danielewicz. It is found that relative source 
functions $S(r)$ have rather similar shapes and zero-separation
intercepts $S(r \rightarrow 0)$. 
Distributions of relative separation have been measured 
out to 20 fm, and the extracted source functions are approximately Gaussian.
We have performed the first experimental check of the predicted connection 
between imaging and traditional meson interferometry techniques and
found that the two methods are in good agreement.  This agreement paves
the way for applications of the imaging method to the interpretation
of pair correlations among strongly interacting particle such as
protons, antiprotons, etc.
Values of source functions at zero separation which are related to the
pion effective volumes of emission are almost constant
across the range of bombarding energies under study.   
\section*{Acknowledgments}
Parts of this paper are based on work done in collaboration with D. Brown and
G.F.~Bertsch~\cite{dbrown_00_1}. 
This research is supported by US DOE, NSF and other funding, as detailed 
in Ref.~\cite{lisa_1,dbrown_00_1}.


\begin{thebibliography}{99}
%%----------------------------experimental details-------

%% E895 beam-time proposal
\bibitem{rai_93}G.~Rai \etal, proposal LBL-PUB-5399 (1993).
%% EOS TPC description
\bibitem{rai_90}G.~Rai \etal, IEEE Trans. Nucl. Sci.  {\bf 37}, 56
(1990).
%% Lisa, pion HBT 
\bibitem{lisa_1}E895 Collaboration, M.A.~Lisa \etal,
Phys. Rev. Lett. {\bf 84}, 2798 (2000).
%% Event mixing ref.
\bibitem{kopylov_74}G. Kopylov, Phys. Lett. {\bf 50B}, 472 (1974).
%% RQMD
\bibitem{sorge_95}H. Sorge, Phys. Rev. C {\bf 52}, 3291 (1995).

%% Brown-Danielevicz method
\bibitem{dbrown_1}D.A.~Brown and P.~Danielewicz,  Phys.~Lett. {\bf{B398}}, 252
(1997).
%% Brown-Danielevicz method
\bibitem{dbrown_2}D.A.~Brown and P.~Danielewicz, Phys.~Rev. {\bf{C57}}, 2474
(1998).
%% Panitkin-Brown paper
\bibitem{panitkin_99_1}S.Y.~Panitkin and D.A.~Brown, Phys.~Rev. {\bf{C61}},
021901 (2000).
%% Panitkin-Brown-Bertsch paper
\bibitem{dbrown_00_1}D.A.~Brown, S.Y.~Panitkin and G.F.~Bertsch,
Phys.~Rev. {\bf{C62}}, 014904 (2000).

%% Inversion reference
\bibitem{tarantola}A. Tarantola, {\em Inverse Problem Theory},
Elsevier, (1987).
%% fourier transforming don't work
\bibitem{FTisbad} D. Brown, nucl-th/9904063.
%%
\bibitem{dbrown_00_2}D.A.~Brown, nucl-th/0003021.
%% imaging using B-splines

\bibitem{dbrown_in_prep}D.A.~Brown and P.~Danielewicz, in preparation;  
C. de Boor, {\em A Practical Guide to Splines}, Springer-Verlag,
(1978).

\end{thebibliography}
\end{document}